# Population Uncertainty in Model Ecosystem: Analysis by Stochastic Differential Equation

Satoru Morita*, Kei-ichi Tainaka, Hiroyasu Nagata and Jin Yoshimura

Department of Systems Engineering, Shizuoka University, Hamamatsu 432-8561

**Abstract:**

Perturbation experiments are carried out by contact process and its mean-field version. Here, the mortality rate is increased or decreased suddenly. It is known that the fluctuation enhancement (FE) occurs after the perturbation, where FE means a population uncertainty. In the present paper, we develop a new theory of stochastic differential equation. The agreement between the theory and the mean-field simulation is almost perfect. This theory enables us to find much stronger FE than reported previously. We discuss the population uncertainty in the recovering process of endangered species.

* morita@sys.eng.shizuoka.ac.jp

Under human managements, ecosystems receive perturbations or stresses, and many species go extinct.[1,2] Recently, coworkers in our laboratory[3,4] studied perturbation experiments to explore the fluctuation enhancement (FE) in simple ecological models. They altered the value of mortality rate $m$ of a species to a higher or lower level; and observed that FE occurred, when the value of $m$ was increased and approached the extinction point ($m_C$). In the present paper, we present a mean-field theory, and apply the mean-field simulation of contact process[5]. It is found that the theory sufficiently predicts the simulation result, and it clarifies the condition of strong FE for the contact process.

We deal with the contact process, which has been extensively studied by various fields, such as mathematics,[5,6] ecology[7,8] and physics.[9,10] We assume that the size ($L$) of a square lattice is finite, but the total number of the lattice sites ($L^2$) is much larger than unity. Each lattice site is labeled by either vacant site (O) or site occupied by the species (X). Interactions are defined by

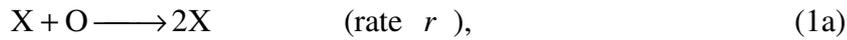

$$X + O \longrightarrow 2X \qquad (\text{rate } r), \qquad (1a)$$

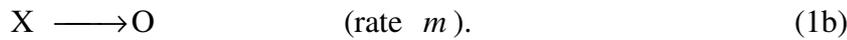

$$X \longrightarrow O \qquad (\text{rate } m). \qquad (1b)$$

The reaction (1a) means the reproduction of the species, and the parameter $r$ is the reproduction rate ($r = 1$ in the simulations). The reaction (1b) and the parameter $m$ denote the death process and the mortality rate, respectively. Simulations are carried not only local interaction (contact process) but also global interactions (mean-field version). We first describe the simulation method for the local interaction.

1) Initially, we distribute individuals X randomly.

2) Spatial pattern is updated in the following two steps:

(i) we perform two-body reaction (1a): Choose one lattice site randomly, and then randomly specify one of four neighboring sites. If the pair of chosen sites are (X,O), then the latter site is changed into X by the rate $r$. If the other pairs are chosen, we skip to the next step.

(ii) we perform one-body reaction (1b). Choose one lattice point randomly; if the site is occupied by X, the site becomes the vacant site (O) by the rate $m$.

3) Repeat step 2) $L^2$ times, where $L^2$ is the total number of lattice sites. This unit of time is called Monte Carlo step[11]

4) The step 3) is repeated.

5) After the system reaches a stationary state, we apply perturbations.

Next, we describe the method for the global interaction in which long-ranged interactions are allowed. The simulation method for the global interaction is almost the same as the local interaction, but the step (i) in 2) for the local interaction is replaced with the following: (i)' Two lattice sites are randomly and independently chosen. If these sites are (X,O), then the latter site is changed into X by rate $r$.

We explain the way of perturbation experiments. Initially (before the perturbation), the mortality rate $m$ is set $m_{BEF}$. After the population reaches a steady state, the value of $m$ is suddenly changed to $m_{AFT}$. We repeat such an experiment $M$ times ($M$ ensembles) with different random numbers. We define the ensemble average $A(t)$ of density and the ensemble variance $V(t)$ as follows:[3,4]

$$A(t) = \frac{1}{M}\sum_{i=1}^{M} x_i(t), \qquad (2)$$

$$V(t) = \frac{1}{M}\sum_{i=1}^{M} [x_i(t) - A(t)]^2, \qquad (3)$$

Here $x_i(t)$ is the density at a moment $t$ in the $i$-th ensemble. The fluctuation enhancement (FE) constitutes a large increase in the value of $V(t)$.

In Fig. 1(a), simulation results for strong FE are displayed. We carry out $M = 100$ repeated experiments for global interaction ($r = 1$ and $L = 100$). Before perturbation ($t < 200$), the mortality rate is set $m_{BEF} = 0.9$. In this case, the steady-state density is low (10%). At the time $t = 200$, the mortality rate is suddenly decreased to $m_{AFT} = 0.3$. In the final state, the density is recovered to 70%. Hence, the perturbation means a recovering process of endangered species. The emergence of such a large FE can be predicted by the following theory.

Now, we derive a stochastic differential equation as a mean-field theory for the case of the global interaction. The population dynamics for the global interaction can be represented by the Chapman-Kolmogorov equation:

$$\begin{aligned} p(x, t + \Delta t) - p(x, t) &= w(x \mid x - \Delta x) p(x - \Delta x, t) \\ &+ w(x \mid \Delta x + 1) p(x + \Delta x, t) \\ &- [w(x + \Delta x \mid x) + w(x + \Delta x \mid x)] p(x, t). \end{aligned} \qquad (4)$$

Here $\Delta t = \Delta x = 1/L^2$, $p(x, t)$ is the probability that the density of the occupied site is $x$ at a moment $t$, and $w(x' \mid x)$ is the transition probability given by

$$w(x + \Delta x \mid x) = r\, x\, (1 - x),$$
$$w(x - \Delta x \mid x) = m\, x.$$

The first line comes from the birth process [reaction (1a)]; the factor $(1-x)$ means the density of empty site. The second line is originated from the death process (1b). When the system size $L^2$ is large, the numerical calculation of (4) is time consuming. Thus, we rewrite the Chapman-Kolmogorov equation to stochastic differential equation as follows.[8,12] In the birth process, the average of the change in the time interval $\Delta t$ is $\Delta x\, w(x + \Delta x \mid x)$, and the variance is $\Delta x^2 [w(x + \Delta x \mid x) - w(x + \Delta x \mid x)^2]$. In the same way, in the death process, the average is $\Delta x\, w(x - \Delta x \mid x)$, and the variance is $\Delta x^2 [w(x - \Delta x \mid x) - w(x - \Delta x \mid x)^2]$. Combining these two processes and neglecting terms of $O(1/L^2)$, we obtain a stochastic differential equation:

$$dx = [rx(1-x) - mx]dt + \frac{\alpha(x)}{L} dW, \qquad (5)$$

where

$$\alpha(x) = \sqrt{rx(1-x) + mx - [rx(1-x)]^2 - (mx)^2}.$$

Here $dW$ denotes Wiener process and is interpreted in Ito's sense. This noise term comes from finite size effect whose variance is inversely proportional to $L^2$. This type of noise is often called demographic fluctuation in ecological science.[13]

Ignoring the noise term, we obtain the deterministic equation of the ensemble average:

$$\frac{dA(t)}{dt} = rA(t)(1 - A(t)) - mA(t). \tag{6}$$

This is just the same as logistic equation, so that there is a stable steady state

$$A_* = 1 - m/r, \tag{7}$$

when the mortality rate $m$ is smaller than a critical value $m_C = r$. When $m > m_C$, extinction occurs immediately.

If demographic noise is small ($L$ is large), the deviation $\xi(t) = x(t) - A(t)$ from the ensemble average $A(t)$ should be small. In the limit of $\xi(t) \to 0$, the dynamics of the deviation vector $\xi(t)$ is given approximately by linear stochastic differential equation (Langevin equation):

$$d\xi(t) = J(t)\ \xi(t)\ dt + \frac{\alpha'(t)}{L} dW. \tag{8}$$

Here $J$ stands for the Jacobian of equations (6). The drift term and the noise term both depend on $t$ through the ensemble average $A(t)$ as follows:

$$\begin{aligned} J(t) &= r - m - 2rA(t), \\ \alpha'(t) &= \sqrt{rA(t)(1-A(t)) + mA(t) - [rA(t)(1-A(t))]^2 - (mA(t))^2}. \end{aligned} \tag{9}$$

Since the variance $V(t)$ is given as the ensemble average of $\xi(t)^2$, we obtain the evolution equation of the variance $V(t)$ as follows:[8,12,14]

$$\frac{d}{dt}V(t) = 2J(t)V(t) + \left(\frac{\alpha'(t)}{L}\right)^2. \qquad (10)$$

Hence, the ensemble average $A(t)$ and the variance $V(t)$ can be estimated by solving equations (6) and (10). The variance in the steady state is obtained as

$$V_* = \frac{m(m^2 + r - mr)}{r^2 L^2}. \qquad (11)$$

The solution (11) is stable, if the following condition holds

$$J_* = r - m - 2rA_* = m - r < 0. \qquad (12)$$

This condition is always satisfied if $m < m_C (= r)$. Equation (11) is an increasing function of $m$ for $r < 3$. Thus, a quasi-static decrease of $m$ causes a decrease of the variance. However, immediately after a rapid change from $m_{BEF}$ to $m_{AFT}$, $J(t) = r - m_{AFT} - 2rA_{*BEF}$, where $A_{*BEF}$ is the species density in the steady state before the perturbation, If $J(t)$ is positive then there is strong FE, because $V(t)$ cannot remains a small value. The condition that $J(t)$ is positive is $m_{AFT} < 2m_{BEF} - r$. By using (7), this condition is rewritten as

$$A_{*AFT} > 2A_{*BEF} \qquad (13)$$

where $A_{*AFT}$ is the species density in the steady state after the perturbation, respectively. This fact indicates that this type of FE can occur when the species density after the perturbation is higher than twice before the perturbation.

We compare the simulation results with the predictions of mean-field theory (6) and (10). We set $r$ to 1 in the simulations. The mortality rate $m$ decreases from $m_{BEF}$ to $m_{AFT}$ ($m_{BEF}$ =0.9 and $m_{AFT}$ =0.3); before perturbation the population size in steady state is set to be small ($A_{*BEF}$ =0.1), but after the perturbation it is significantly large ($A_{*AFT}$ =0.7). This setting follows the condition (13). In Fig. 1, the results of such an experiment are displayed, where the upper and lower figures represent the simulation for global interaction and the mean-field theory, respectively. From Fig. 1, we find that there is a sharp peak of $V(t)$ both in the simulation and the theoretical result. The theory indicates that this sharp peak is caused by the unstable time interval of $J(t) > 0$. The similar result is obtained for local interaction (Fig. 2). We use $m_{BEF}$ =0.59 and $m_{AFT}$ =0.27 in order to change the steady-state density from 10% and 70%. From Fig. 2(b), we find the FE for local interaction is higher and wider than that for global interaction. This is because the relaxation time for the local interaction is longer, so that the unstable time interval should be longer.

Fig. 3 displays the case of the global interaction for $m_{BEF}$ =0.3 and $m_{AFT}$ =0.9. This case is just opposite to the case in Fig. 1. This figure shows that the variance increases monotonously and approaches a new steady state, which is calculated by (11). There is not FE because $J(t)$ is always negative. The similar result is obtained for the local interaction. Fig. 4 displays the result of the local interaction ($m_{BEF}$ =0.27 and $m_{AFT}$ =0.59). For the local interaction, the increase of the variance is larger but slower than the global interaction. The reason is again because of a longer relaxation time for the local interaction. For this parameter, the qualitatively same phenomena are observed in both global and local cases. However, coworkers in our laboratory reported that when $m_{AFT}$ is near and above $m_C$ the large but slow FE occurred for local interaction, but was never observed for global interaction.[3] This phenomena are caused by extinction, which is not taken consider in our theory. For local interaction, extinction takes a much longer time than for global interaction. This type of FE may be come from spatial effects.[3,4]

In summary, we developed the mean-field theory by using stochastic differential equation. We derived ordinary differential equations for both ensemble average (6) and variance (10). This theory enables us to predict the magnitude of FE. Here we report two cases: the recovering process of low-density species and the opposite process. In the former a large FE can emerge for both local and global interactions. For global interactions, the condition that FE emerges is given by (13). The agreement between the global simulation and the mean-field theory is almost perfect. In the latter, no FE occurs for neither local nor global interactions, so long as $m_{AFT} < m_C$. However, if $m_{AFT} \cong m_C$, the extinction occurs, so that FE was observed for local interaction but not for global interaction. Our results suggest that when an endangered species is recovered, strong FE emerges. The magnitude of FE is much larger, compared to the previous results.[3]

Finally, the ecological meaning of our results is discussed. We report the population uncertainty in the recovering process of endangered species. This means that the quantitative forecast is essentially difficult after a conservation policy. Sometimes the recovering process proceeds very slowly, but sometimes the endangered species is immediately recovered.

**ACNOWLEGEMENT**

The authors wish to thank Mr. Taro Hayashi for his collaboration. S.M. is supported by Grants-in-Aid for Scientific Research from the Japan Society of Promotion of Science under the contract number CE19740234.

process (Poisson process) where the squared terms $[rx(1-x)]^2 - (mx)^2$ are removed: $\alpha = \sqrt{rx(1-x) + mx}$.[14,15] Our updating steps (i) and (ii) in 2) are not Poissonian for simplicity of simulation. This difference is not important for FE.

[14] L.J.S Allen: *An introduction to stochastic processes with applications to biology* (Pearson Education Inc., New Jersey 2003).

[15] C.W. Gardiner: *Handbook of stochastic methods for physics, chemistry and the natural sciences* (Spinger-Verlag, Berlin Heidelberg New York 2004).

**Figure captions**

**Figure 1.** Typical results of FE for a recovering process ($r=1$ and $L=100$). (a) simulation results, (b) theoretical prediction. Before perturbation ($t<200$), the mortality rate is set $m_{BEF}$ =0.9; the steady-state density is low (10%). At the time $t=200$, the rate is suddenly decreased to $m_{AFT}$ =0.3 (70% density).

**Figure 2.** Same as Fig. 1, but for local interaction ($m_{BEF}$ =0.59 and $m_{BEF}$ =0.27).

**Figure 3.** Typical results of perturbation experiment. The change of mortality rate is just opposite to the case in Fig. 1 ($m_{BEF}$ =0.3 and $m_{AFT}$ =0.9).

**Figure 4.** Same as Fig. 3, but for local interaction ($m_{BEF}$ =0.27 and $m_{BEF}$ =0.59).

Figure 1

(a) Result of global simulation

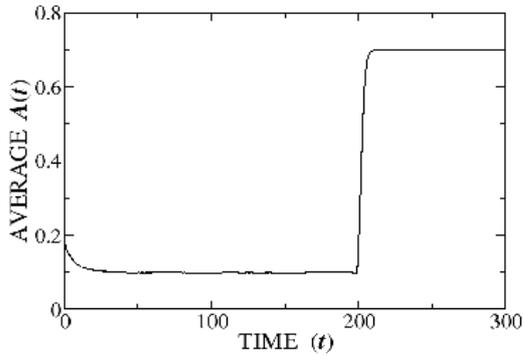
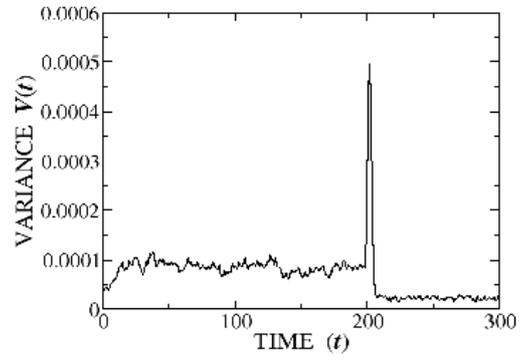

(b) Prediction by mean-field theory

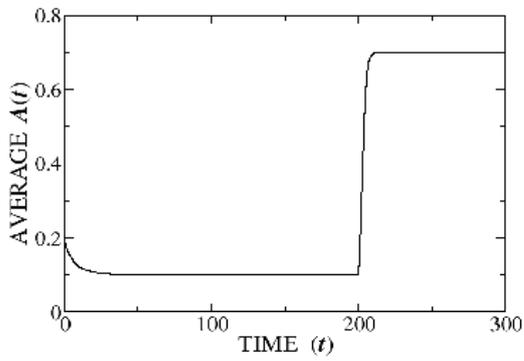
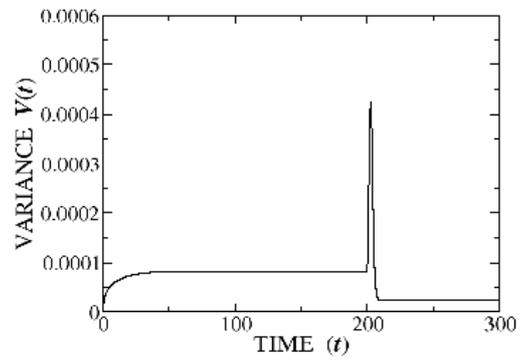

Figure 2

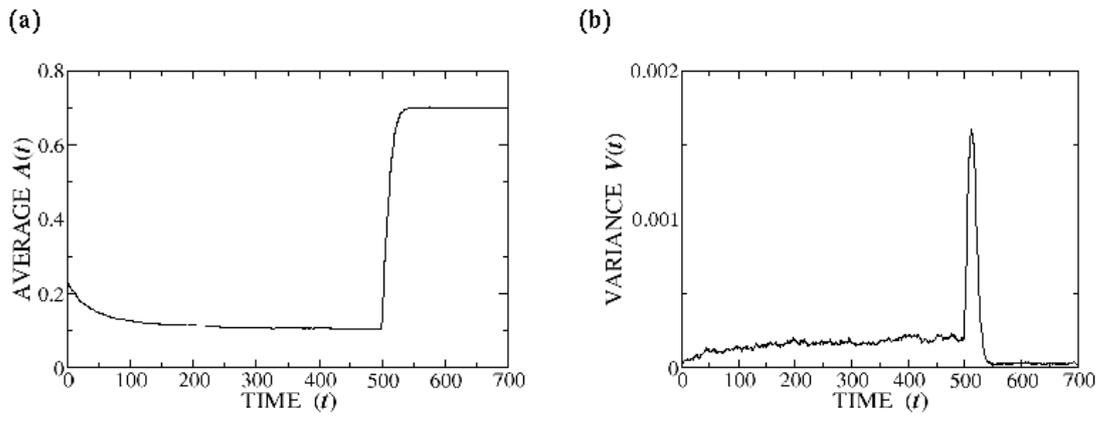

Figure 3

Result of global simulation

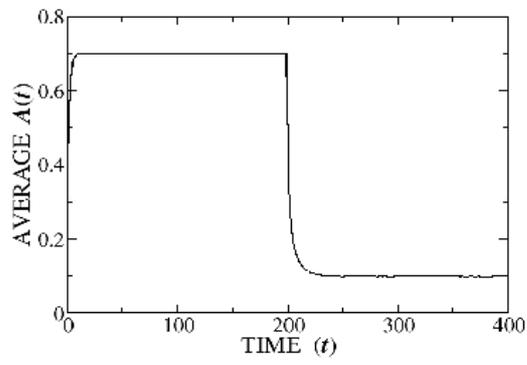
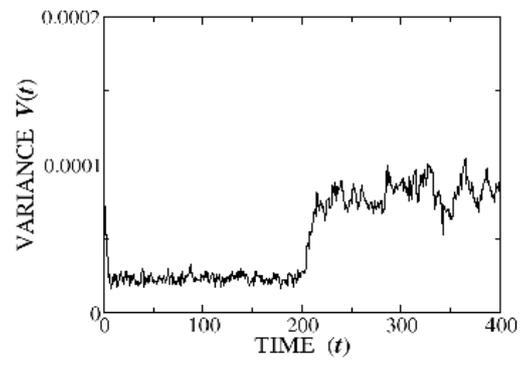

Prediction by mean-field theory

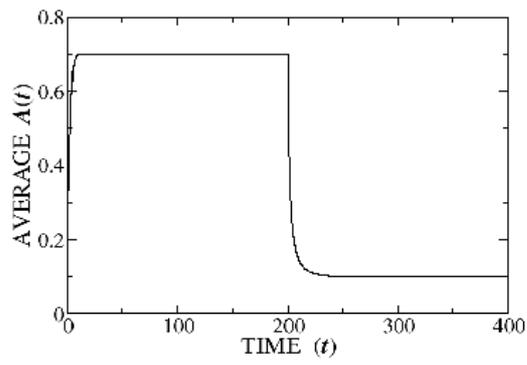
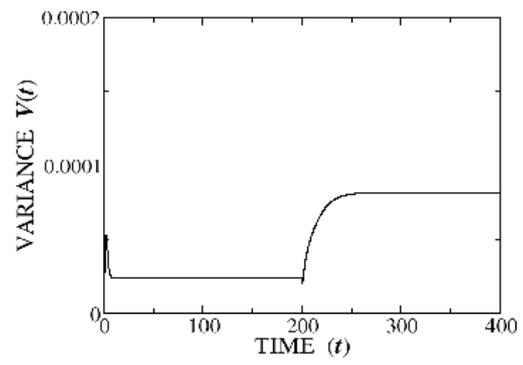

Figure 4

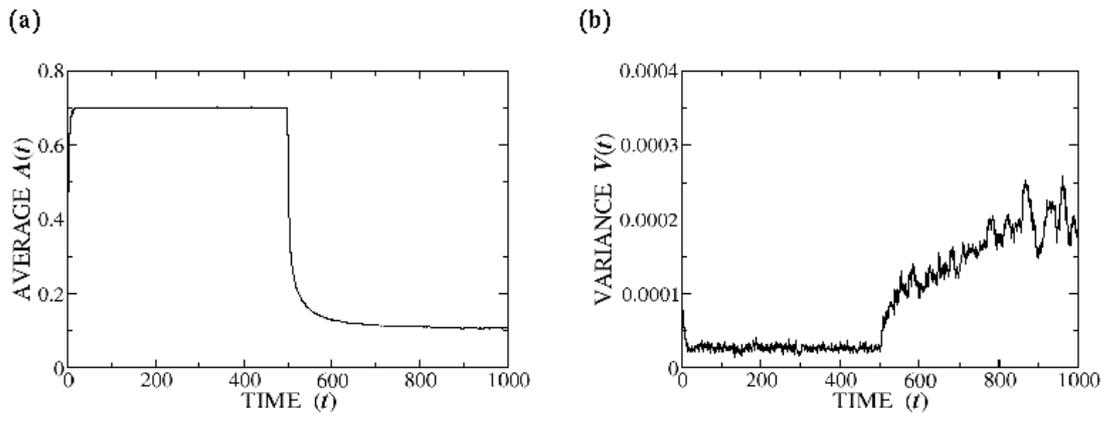